\documentclass[usenatbib]{mn2e}
\usepackage[centering,totalwidth=480pt,totalheight=680pt]{geometry}
\usepackage{psfig}
\usepackage[dvips]{graphicx}

\usepackage{amsmath}

\newcommand{\msun}{\mbox{${\rm M}_{\odot}$}}
\newcommand{\lsun}{\mbox{${\rm L}_{\odot}$}}
\newcommand{\mbh}{\mbox{$m_{\rm BH}$}}

\newcommand{\mgal}{\mbox{$m_{\rm gal}$}}
\newcommand{\msph}{\mbox{$m_{\rm sph}$}}
\newcommand{\sigmabh}{\mbox{$\sigma_{\rm BH}$}}
\newcommand{\fagn}{\mbox{$f_{\rm AGN}$}}
\newcommand{\fedd}{\mbox{$f_{\rm Edd}$}}

\title[Evolution of the \mbh-\mgal\ Relationship]{Empirical Constraints
  on the Evolution of the Relationship between Black Hole and Galaxy
  Mass: Scatter Matters}

\author[R. S. Somerville] {
\parbox[t]{\textwidth}{ 
Rachel S. Somerville$^{1,2}$\thanks{E-mail: somerville@stsci.edu}
}
\vspace*{6pt} \\
$^1$ Space Telescope Science Institute, 3700 San Martin Dr., Baltimore, MD 21218\\
$^2$ Department of Physics and Astronomy, Johns Hopkins University, Baltimore, MD 21218\\
}

\begin{document}

\newcommand{\refs}{{\bf (ref)}}
\newcommand{\notice}[1]{{\it [{#1}]}}
\newcommand{\plotone}[1]
           {\centering \leavevmode \psfig{file=#1,width=\columnwidth,clip=}}
\newcommand{\plotside}[1]
           {\centering \leavevmode \psfig{file=#1,width=\textwidth,clip=}}
\newcommand{\plottwo}[2]
           {\centering \leavevmode \psfig{file=#1,width=\columnwidth,clip=}
                            \hfill \psfig{file=#2,width=\columnwidth,clip=}}
\newcommand{\plotfull}[1]
           {\centering \leavevmode \psfig{file=#1,width=\textwidth,clip=}}
\newcommand{\plotwhole}[1]
           {\centering \leavevmode
             \psfig{file=#1,width=\textwidth,height=8.2in,clip=}}

\def\lesssim{\lower.5ex\hbox{$\; \buildrel < \over \sim \;$}}
\def\gtrsim{\lower.5ex\hbox{$\; \buildrel > \over \sim \;$}}

\maketitle

\begin{abstract}

I investigate whether useful constraints on the evolution of the
relationship between galaxy mass (\mgal) and black hole (BH) mass
(\mbh) can be obtained from recent measurements of galaxy stellar mass
functions and QSO bolometric luminosity functions at high redshift. I
assume a simple power-law relationship between \mgal\ and \mbh, as
implied by BH mass measurements at low redshift, and consider only
evolution in the zero-point of the relation. I argue that one can
obtain a lower limit on the zero-point evolution by assuming that
every galaxy hosts a BH, shining at its Eddington rate. One can obtain
an upper limit by requiring that the number of massive BH at high
redshift does not exceed that observed locally. I find that, under
these assumptions, and neglecting scatter in the \mgal-\mbh\ relation,
BH must have been a factor of $\sim 2$ larger at $z\sim1$ and 5--6
times more massive relative to their host galaxies at
$z\sim2$. However, accounting for intrinsic scatter in
\mgal-\mbh\ considerably relaxes these constraints. With a logarithmic
scatter of 0.3--0.5 dex in \mbh\ at fixed \mgal, similar to estimates
of the intrinsic scatter in the observed relation today, there are
enough massive BH to produce the observed population of luminous QSOs
at $z\sim2$ even in the absence of any zero-point evolution. Adopting
more realistic estimates for the fraction of galaxies that host active
BH and the Eddington ratios of the associated quasars, I find that the
zero-point of the \mgal-\mbh\ relation at $z\sim2$ cannot be much more
than a factor of two times larger than the present-day value, as the
number of luminous quasars predicted would exceed the observed
population.

\end{abstract}

\begin{keywords}

black hole physics -- galaxies: formation -- galaxies: evolution --
quasars: general -- galaxies: active -- cosmology:theory

\end{keywords}

%=======================
% 1
\section{Introduction}
\label{sec:intro}
%=======================

The discovery of the relationship between the masses of nuclear
supermassive black holes (SMBH) and the luminosity, velocity
dispersion, or mass of their host galactic spheroids
\citep{dressler:89,kormendy_review:95,magorrian:98,ferrarese:00,gebhardt:00,marconi:03,haering:04}
is surely one of the most profound observational results of the past
decade, if not the past century. Different methods applied to both
dormant and active BH in the nearby Universe now yield consistent
results and indicate that BH mass and galaxy velocity dispersion
$\sigma$ are related via $\mbh \propto \sigma^\beta$, where $\beta
\simeq 4$--5 \citep{ferrarese:00,gebhardt:00,tremaine:02}, and BH mass
and galaxy mass are related via $\mbh \propto \mgal^{1.1}$
\citep{marconi:03,haering:04}\footnote{Several other equally tight
  relationships between BH mass and other galaxy properties have been
  discovered \citep[e.g.][]{graham:01,kormendy:09}. Although these
  relationships are intriguing, here I focus on the relationship
  between BH mass and galaxy mass.}. The {\em observed} scatter in
both of these relationships is remarkably small, and implies intrinsic
scatters of approximately 0.3 dex in $\mbh$ at fixed $\sigma$, and
0.5 dex in $\mbh$ at fixed $L$ \citep{novak:06,gultekin:09}.

A large number of theoretical explanations for the origin of this
observed relationship (hereafter referred to for brevity as the
\mbh-\mgal\ relationship) have been proposed
\citep[e.g.][]{silk_rees:98,burkert_silk:01,adams:01,adams:03,wyithe_loeb:03,robertson:06,croton:bhev,hopkins_bhfpth:07}.
However, there is no widely accepted unique theoretical model, and the
models differ in their predictions for the amount of evolution in the
\mbh-\mgal\ relationship.  This quantity is therefore a potentially
strong discriminator between different theoretical models, and there
is great interest in obtaining robust direct observational
measurements of this relationship at high redshifts. A great deal of
effort and telescope time has been expended towards achieving this
goal.

With currently available facilities, the masses of dormant BH can be
measured via the dynamics of the surrounding gas or stars
\citep[see][for a recent review]{ferrarese_ford:05} only for very
nearby galaxies. At high redshift, it is currently possible to attempt
to measure masses only for active or accreting BH. Several such
studies have claimed to find evidence for significant evolution in the
\mbh-\mgal\ relationship, always in the sense that black holes are
more massive at high redshift relative to their host galaxy
\citep[e.g.][]{treu:04,peng:06,woo:06,woo:08,salviander:07}.  However,
these methods rely on a set of simplified underlying assumptions and
various proxies for the desired quantities, and are subject to
potentially severe selection biases.  \citet{lauer:07} have argued
that if there is a moderate amount of scatter in the intrinsic
\mbh-\mgal\ relationship (consistent with observational constraints on
the scatter in the local relation), these selection biases can account
for most or all of the claimed evolution. It is therefore interesting
to explore the possibility of obtaining independent {\em empirical}
constraints on the evolution. Even obtaining robust upper and lower
limits on the evolution could be useful.

Recently, it has become common practice to estimate the stellar masses
of galaxies from multi-wavelength broadband photometry and/or
spectroscopy \citep[e.g.][]{bell_dejong:01,bell:03,kauffmann:03}. The
availability of deep multi-wavelength surveys covering substantial sky
area has yielded estimates of the stellar mass functions out to high
redshift, $z\sim 4$--5
\citep[e.g.][]{Drory:04,Fontana:04,Borch:06,Fontana:06,Pannella:06,Bundy:06,Pozzetti:07,Vergani:08,Marchesini:08,PerezGonzalez:08}. Similarly,
there has been significant progress in piecing together the evolution
of the {\em bolometric} luminosity function of quasars and Active
Galactic Nuclei (AGN) out to high redshift ($z\sim 6$) from
multi-wavelength surveys \citep[][and references therein; hereafter
  HRH07]{hopkins_qsolf:07}.

In this paper, I explore whether one can derive useful empirical
constraints on the evolution of the relationship between BH mass and
galaxy mass by comparing these two sets of observed statistical
distributions (galaxy stellar mass functions and QSO/AGN luminosity
functions) under the basic and well-accepted ansatz that accreting BH
provide the power source for AGN. The outline of the rest of the paper
is as follows. In \S\ref{sec:results}, I describe my basic set of
assumptions. In \S\ref{sec:results:noscat} I present upper and lower
limits on the evolution of the \mbh-\mgal\ relation, assuming that
there is no intrinsic scatter in the relation. In
\S\ref{sec:results:scat} I discuss how these constraints are impacted
by the inclusion of intrinsic scatter. I conclude in
\S\ref{sec:conclusions}.

\begin{figure*} 
\begin{center}
\plottwo{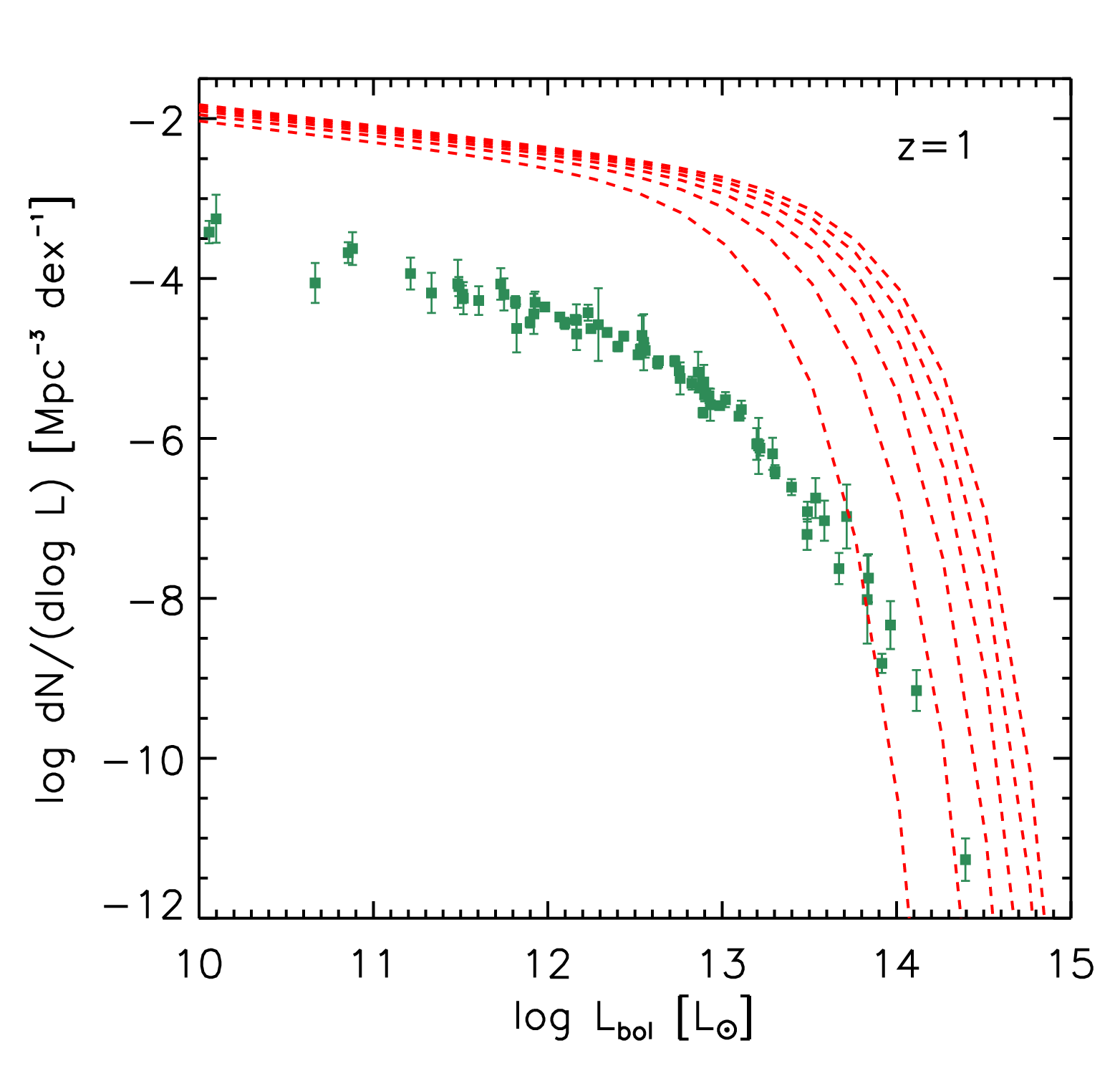}{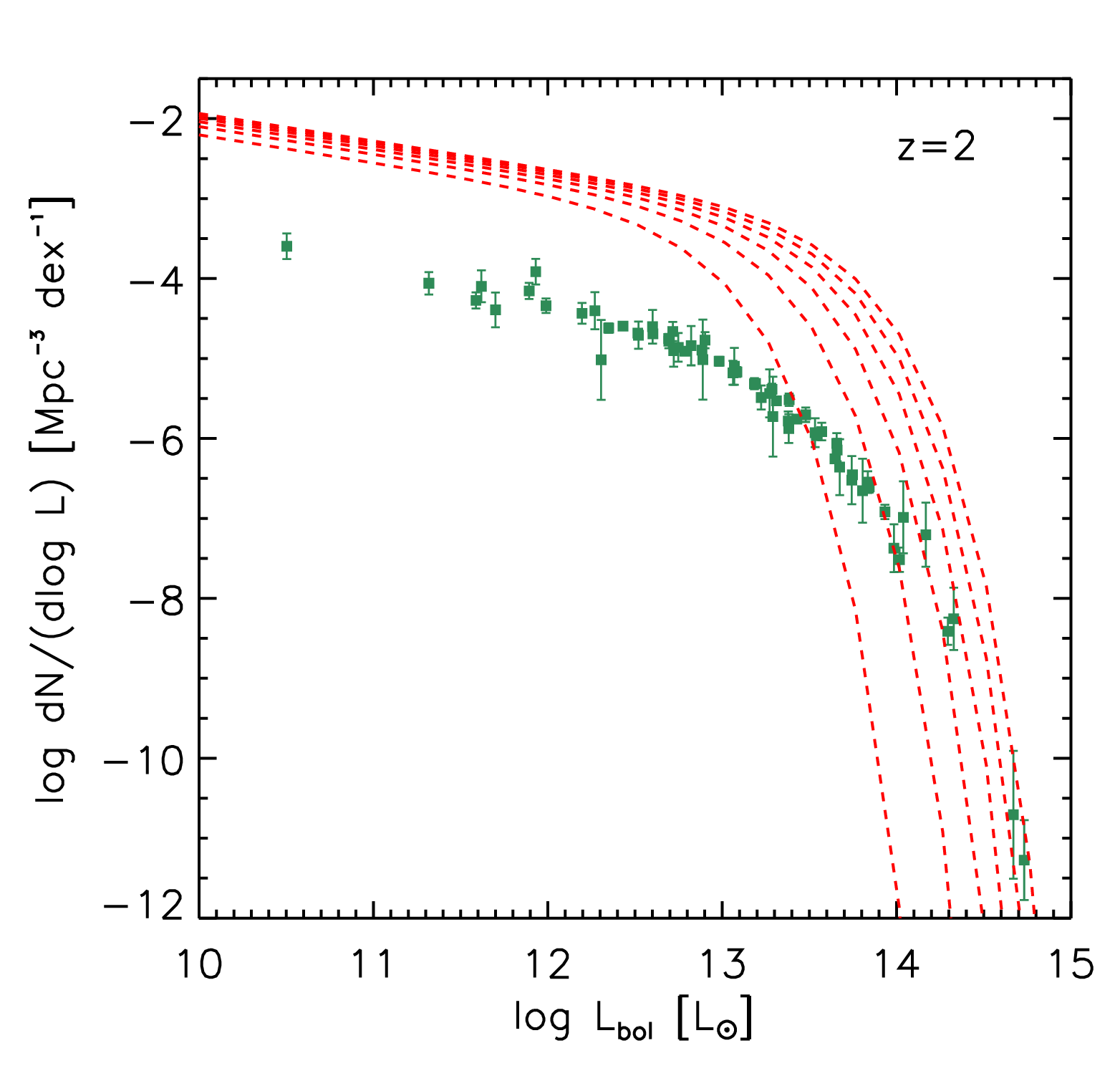}
\end{center}
\caption{\small The bolometric luminosity function of quasars at $z=1$
  (left panel) and $z=2$ (right panel). Square symbols with error bars
  show the estimate of the observed bolometric QSO LF from HRH07. Dashed
  lines show the upper limit on the QSO LF derived from the observed
  stellar mass function at the relevant redshift and the arguments
  described in the text. From left to right, the dashed lines assume
  that the zero-point of the \mbh-\mgal\ relation has evolved by a
  factor of $\Gamma=1$, 2, 3, 4, 5, or 6. Under these assumptions, the
  mass of a typical BH hosted by a galaxy of a given mass must have
  been larger by a factor of $\sim 2$ at $z=1$ and by a factor of 5--6
  at $z=2$.
\label{fig:qsolf}} 
\end{figure*}

%=======================
% 2
\section{Upper and Lower Limits on the Evolution}
\label{sec:results}
%=======================

For nearby dormant BH, the average relationship between BH mass and
galaxy mass can be characterized as $\mbh \propto \mgal^{1.1}$
\citep{marconi:03,haering:04}\footnote{\citet{lauer:07a} suggest a
  slope of unity, but this would not significantly effect the
  results.}. I will explore the simplest possible form for possible
evolution of this relationship, namely scaling by a factor that is a
function of redshift only $\Gamma(z)$. Thus the mass of a BH residing
in a galaxy with mass \mgal\ at redshift $z$ is given by:
\begin{equation}
\mbh(z,\mgal) = \Gamma(z)\, \mbh(z=0, \mgal) = \Gamma(z)\, \mgal^{1.1} \, .
\label{eqn:mbhev}
\end{equation}

I now make use of the observed galaxy stellar mass function at some
redshift of interest, and assume that every galaxy hosts a SMBH with
mass given by Eqn.~\ref{eqn:mbhev}. One can then obtain a reasonable
{\em lower} limit on the evolution in \mbh-\mgal\ (i.e. on
$\Gamma(z)$) at a given redshift by assuming that 1) every BH is
active at all times (has a duty cycle of unity) 2) every active BH
always radiates at its Eddington luminosity. This set of assumptions
will clearly maximize the number of luminous quasars for a given
population of BH, under the fairly standard conjecture of
Eddington-limited accretion.

One can then obtain an {\em upper} limit on the evolution by comparing
the implied BH mass function at the redshift of interest with
observational estimates of the {\em present-day} BH mass
function. Again, under the apparently reasonable assumptions that BH
masses increase monotonically with time and that significant numbers
of massive BH are not somehow lost from galaxies, clearly the number
of massive BH in the past cannot exceed that at the present day. To
state the condition more precisely, the number density of BH,
$\phi(\mbh)$, at high redshift may not exceed the present day value
for all BH masses greater than some threshold value $\mbh > M_{\rm
  min}$.

For the observed stellar mass function as a function of redshift, I
adopt the fitting functions of \citet{Fontana:06}, based on
measurements from the GOODS-MUSIC survey. I have checked that these
fitting functions provide good agreement with the results from other
surveys (see \citet{fontanot:09} for a thorough comparison of stellar
mass function estimates from different surveys over a wide range of
redshifts). There is good agreement between different estimates of the
stellar mass function up to $z\sim 2$; at higher redshifts the results
of different studies diverge. For this reason, the results presented
here are limited to redshift two and below. The stellar masses have
been converted to correspond to a \citet{chabrier:03} stellar initial
mass function. I will assume that there are no significant
redshift-dependent systematic errors in the stellar mass
estimates. For the moment, because I am mainly exploring the
feasibility of this approach, I also ignore the effect of the random
errors on the stellar masses, although these will probably have an
impact on the quantitative results.

\subsection{Constraints without Scatter}
\label{sec:results:noscat}

Initially, I assume that the relationship between BH mass and galaxy
mass has no intrinsic scatter. It is then straightforward to derive
the implied bolometric luminosity function of QSO/AGN from the
observed galaxy stellar mass function under the set of assumptions
outlined above, for a given value of the evolution factor
$\Gamma(z)$. Figure~\ref{fig:qsolf} shows the observed bolometric QSO
luminosity function as estimated by HRH07 along with the upper limit
estimate based on the stellar mass function, for different values of
$\Gamma$. This comparison is shown at $z=1$ and $z=2$. At QSO
luminosities below the ``knee'' in the LF, the upper limit estimate
overproduces QSOs, which can be understood as implying that these BH
are not active at all times and/or are radiating at sub-Eddington
luminosities. What is more interesting is that at the highest
luminosities, above $\sim 10^{14} \lsun$ at $z=1$ or $10^{13.5}$ at
$z=2$, {\em without evolution in the \mbh-\mgal\ relation, the number
  of luminous QSOs is significantly underestimated}. Assuming that
these luminous QSOs are not radiating above their Eddington
luminosity, and are not magnified somehow (e.g. by beaming or
lensing), one can then read off the {\em minimum} amount of evolution
(minimum value of $\Gamma$) required to produce enough luminous
QSOs. This corresponds to $\Gamma_{\rm min} \sim 2$ at $z=1$ and
$\Gamma_{\rm min} \sim 5$--6 at $z=2$. Taken at face value, then,
these results require that BH were 5--6 times larger for a given
galaxy mass at $z=2$.

\begin{figure*} 
\begin{center}
\plottwo{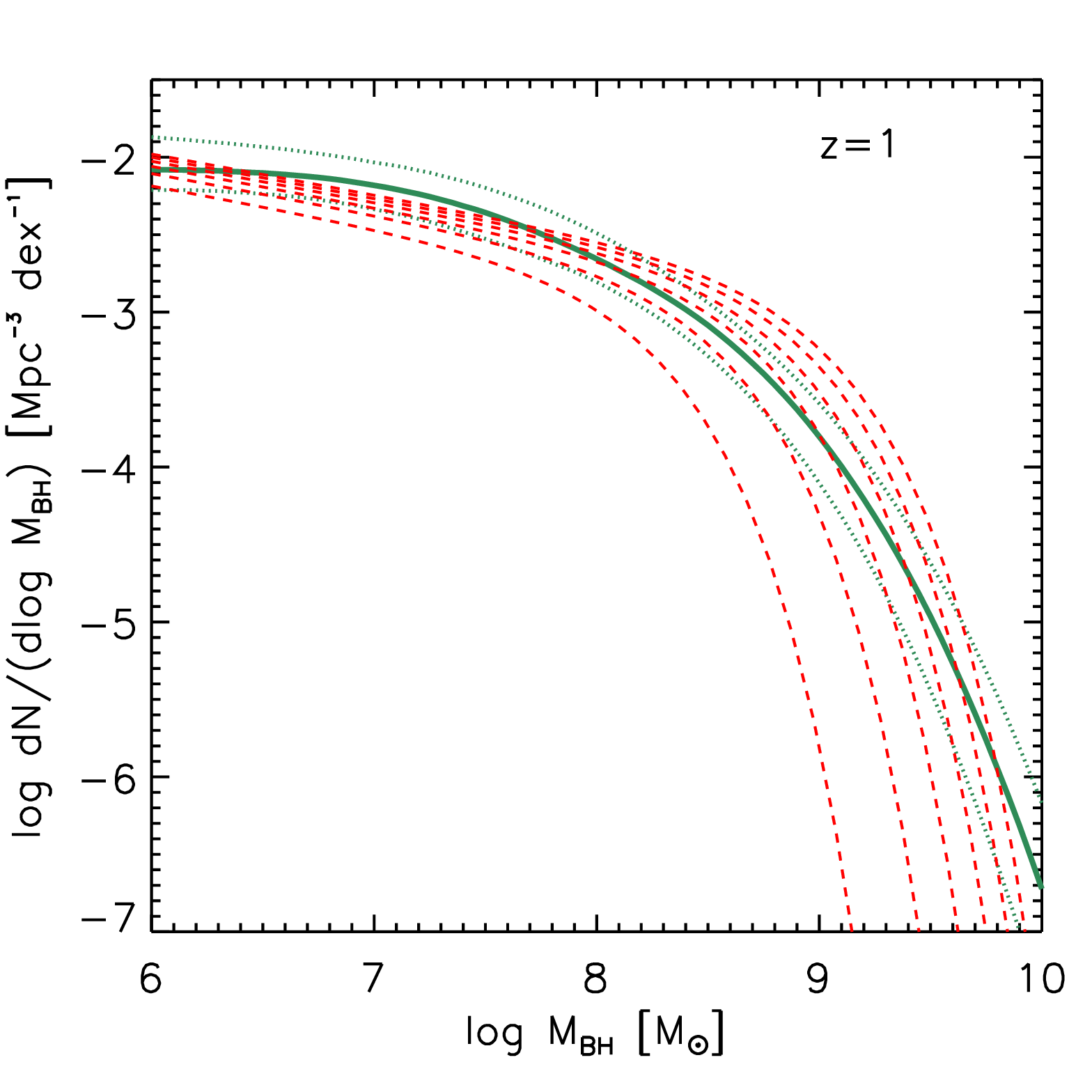}{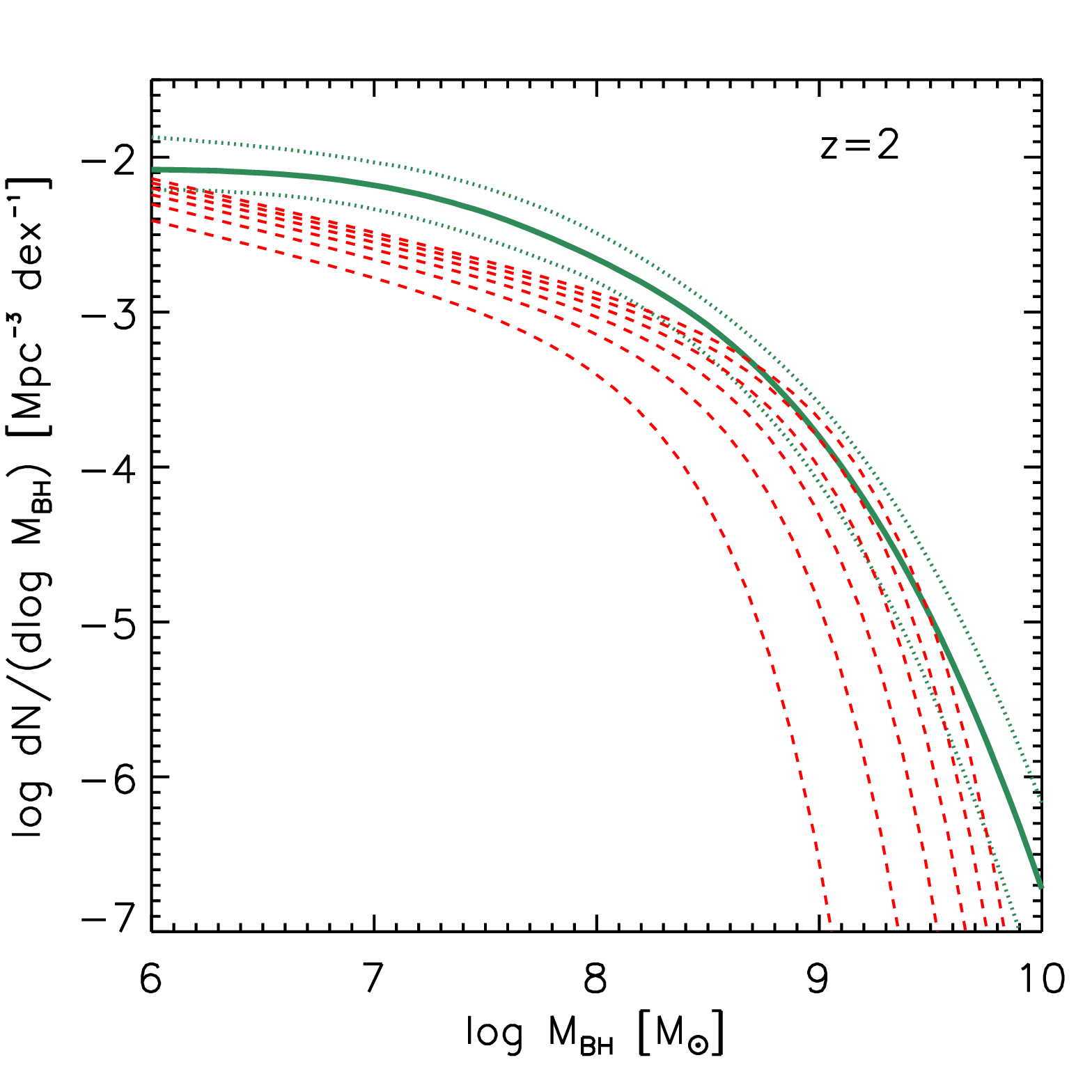}
\end{center}
\caption{\small The mass function of SMBH. The solid green line shows
  the observational estimate of the BH mass function at $z=0$ from
  \protect\citet{marconi:04}. Dashed (red) lines show the BH MF
  implied by the observed galaxy stellar mass function at $z=1$ (left
  panel) or $z=2$ (right panel), the relationship between BH mass and
  galaxy mass described in the text, and evolution in the zero-point
  of the \mbh-\mgal\ relation of a factor of $\Gamma=1$, 2, 3, 4, 5,
  or 6 (curves from left to right, respectively).
\label{fig:bhmf}}
\end{figure*}

Now let us consider the upper limit on the evolution, or maximum
allowed value of $\Gamma$. Figure~\ref{fig:bhmf} shows the
observational estimate of the BH mass function\footnote{Note that the
  \protect\citet{marconi:04} estimate of the BHMF includes the effect
  of scatter in the local \mbh-\mgal\ relation, and therefore the
  comparison shown here is not strictly self-consistent.  However, I
  will consider scatter self-consistently in the next Section.} at
$z=0$ \citep{marconi:04}, compared with the results from the stellar
mass function at $z=2$ scaled by the same series of values of
$\Gamma$. Clearly, as long as BH cannot decrease in mass or be ejected
from their host galaxies, then there is an upper limit of $\Gamma_{\rm
  max} \sim 6$ at $z\sim2$. It is interesting that the lower limit
from the QSO LF, discussed above, and this upper limit are so close to
one another.

\subsection{Constraints with Scatter}
\label{sec:results:scat}

\begin{figure*} 
\begin{center}
\includegraphics[width=6.5in]{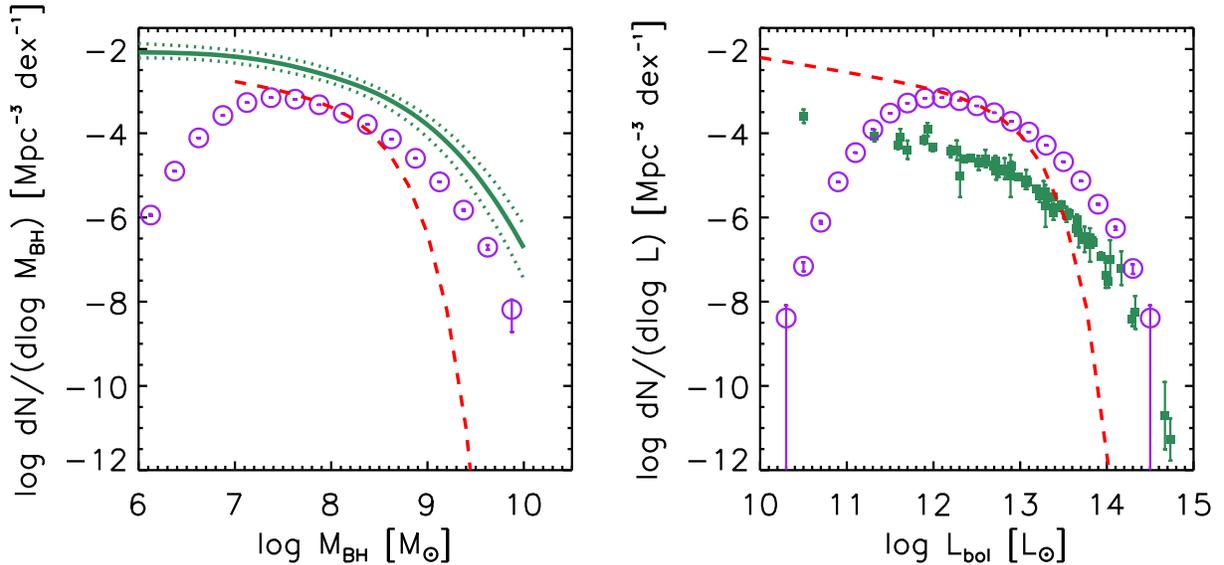}
\end{center}
\caption{\small Left: BH mass function. Open (purple) dots show the
  BHMF implied by the observed $z=2$ GSMF, no evolution in the
  \mbh-\mgal\ relation ($\Gamma=1$), and a scatter of $\sigmabh=0.3$;
  error bars are simple Poisson errors. The solid (green) line shows
  the observational estimate of the BH mass function at $z=0$ from
  \protect\citet{marconi:04}; the dashed (red) line shows the BH MF
  implied by the observed galaxy stellar mass function at $z=2$ plus
  the assumed \mbh-\mgal\ relation with no evolution ($\Gamma=1$) and
  no scatter. To satisfy the constraint, the purple dots
    (prediction) should be {\em lower} than the green line
    (observations). Right: The QSO luminosity function. Open (purple)
  dots show the upper limit on the QSO LF from the same argument, but
  including scatter in the \mbh-\mgal\ relation ($\sigmabh=0.3$);
  error bars are Poisson. Square (green) symbols with error bars show
  the estimate of the observed bolometric QSO LF at $z=2$ from HRH07;
  the dashed (red) line shows the upper limit on the QSO LF derived
  from the observed GSMF at $z=2$ and the arguments described in the
  text, for $\Gamma=1$ and $\sigmabh=0$.  Here, to satisfy the
    constraint, the purple dots (prediction) should be {\em higher}
    than the green squares (observations). The inclusion of a
  moderate amount of scatter has a large impact on the high-mass end
  of the BHMF and the high luminosity end of the QSO LF. When
    scatter is included, the assumption that the \mbh-\mgal\ relation
    has not evolved since $z\sim2$ appears to be consistent with these
    constraints. 
\label{fig:bhev_s0.3}}
\end{figure*}

\begin{figure*} 
\begin{center}
\includegraphics[width=6.5in]{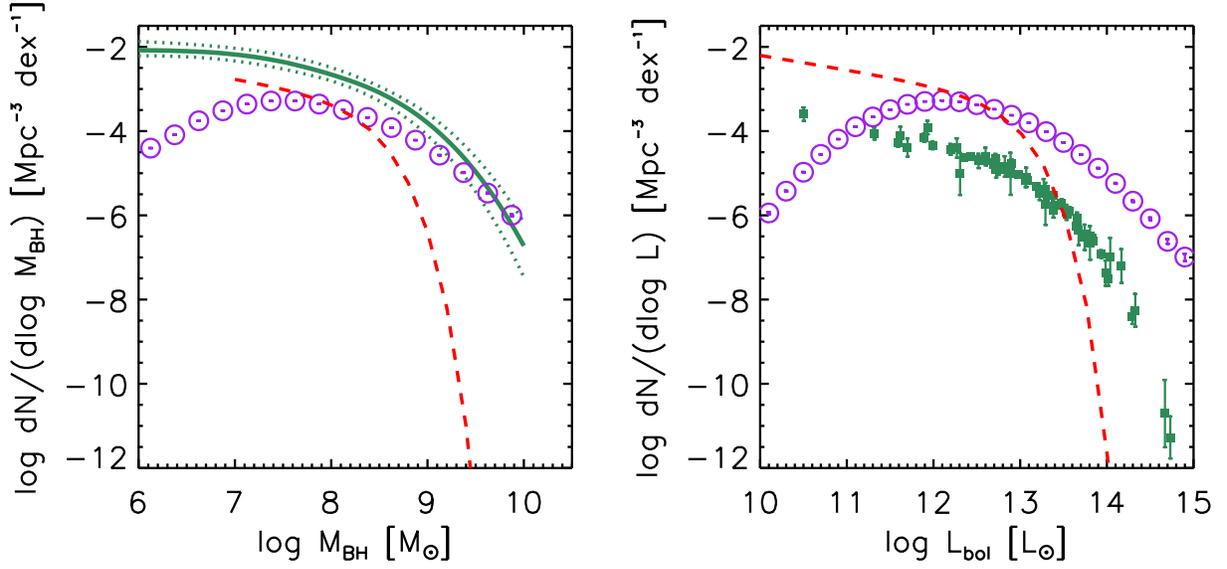}
\end{center}
\caption{\small The same as Fig.~\protect\ref{fig:bhev_s0.3}, but with
  $\sigmabh=0.5$.
\label{fig:bhev_s0.5}}
\end{figure*}

\begin{figure*} 
\begin{center}
\includegraphics[width=6.5in]{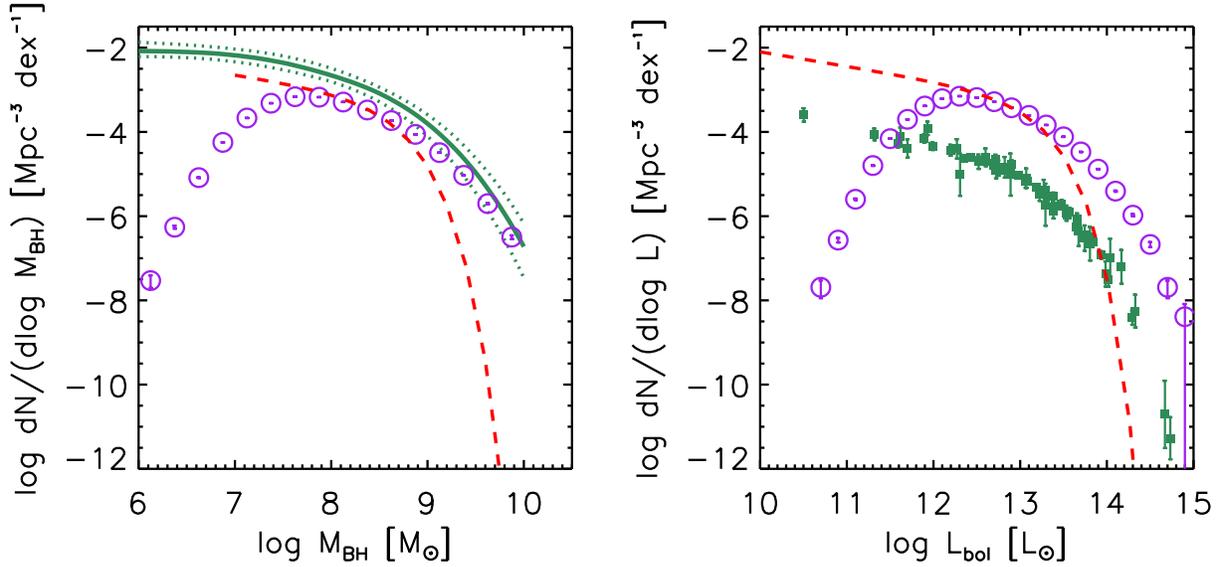}
\end{center}
\caption{\small The same as Fig.~\protect\ref{fig:bhev_s0.3}, but with
  $\Gamma=2$ and $\sigmabh=0.3$.
\label{fig:bhev_g2}}
\end{figure*}

So far I have neglected intrinsic scatter in the
\mbh-\mgal\ relation. However, as noted by \citet{lauer:07}, because
of the very steep decline of the high-mass end of the galaxy mass or
luminosity function, the highest mass black holes are actually more
likely to be outliers from the \mbh-\mgal\ relation, hosted by galaxies
of more modest mass, rather than typical BH in the much rarer
high-mass galaxies that would be the sole hosts of such BH in the
absence of scatter.

\citet{novak:06} attempted to constrain the {\em intrinsic} scatter in
the $\mbh-\sigma$ and $\mbh-L$ relations at $z=0$, and concluded that
due to the small sample of galaxies with reliable measurements of BH
mass and uncertainties in the observational error estimates on \mbh,
only upper limits on the scatter could be obtained. They estimated
these upper limits to be $\delta_{\sigma} < 0.3$ and $\delta_L < 0.5$,
where $\delta_{\sigma}$ is the 1-$\sigma$ log scatter in the
$\mbh-\sigma$ relation and $\delta_L$ is the same for the $\mbh-L$
relation. \citet{marconi:03} find a similar scatter in the
$\mbh-\msph$ relation as in $\mbh-\sigma$, and \citet{haering:04}
find an {\em observed} scatter of 0.3 dex in the $\mbh-\mgal$
relation, implying that the intrinsic scatter is presumably
smaller. Recently, \citet{gultekin:09} made a detailed study of the
magnitude of the intrinsic scatter in $\mbh-\sigma$ and $\mbh-L$,
finding $\delta_{\sigma} = 0.31$ for ellipticals but a larger scatter
of $\delta_{\sigma} = 0.44$ for all galaxies (including spirals). They
furthermore find that the shape of the distribution of the intrinsic
residuals in \mbh\ at fixed $\sigma$ is consistent with a log-normal
(and inconsistent with a normal distribution).
 
Unfortunately, almost nothing is known about the possible evolution of
the intrinsic scatter in the BH-galaxy scaling relations. Therefore, I
investigate how the inclusion of representative amounts of scatter
would impact the results presented in
Section~\ref{sec:results:noscat}. In order to do this, I run Monte
Carlo simulations of $\sim 10^6$ galaxies, in which I first select
galaxy masses from the observed stellar mass function at the redshift
of interest (using $z=2$ as a representative case). I then assign BH
masses to each galaxy according to Eqn.~\ref{eqn:mbhev}, adding a
random deviate in mass selected from a log-normal distribution with
root variance $\sigmabh$, and consider the implied BH mass function
and upper limit on the QSO LF as before.

Results of these experiments for various values of $\sigmabh$ and
$\Gamma$ (all at $z=2$) are shown in
Figures~\ref{fig:bhev_s0.3}--\ref{fig:bhev_g2}. Note that I cut off
the galaxy stellar mass function below $10^{10} \msun$ because these
low-mass galaxies do not provide interesting constraints and including
them causes the Monte Carlo simulations to take much longer to run
(for a given desired number of high-mass objects). In
Figure~\ref{fig:bhev_s0.3}, one can see that when a moderate scatter
in the \mbh-\mgal\ relation is included ($\sigma_{BH} = 0.3$, similar
to the scatter in the observed relation at the present day), the
number of luminous QSOs can be reproduced under the fairly extreme
assumptions used in the lower limit exercise (all BH radiate at their
Eddington limit at all times). Even a scatter half as large as the
observed present-day estimates ($\sigma_{BH} = 0.15$), with no
evolution in the normalization ($\Gamma=1$) marginally satisfies the
lower limit. With a slightly larger scatter ($\sigma_{BH} = 0.5$) or
moderate evolution in the zero-point $\Gamma=2$ (see
Fig.~\ref{fig:bhev_s0.5} and \ref{fig:bhev_g2}), bright QSOs are
overproduced by a factor of 10-100, leaving room for more reasonable
assumptions about duty cycle and Eddington ratio.

\begin{figure} 
\begin{center}
\plotone{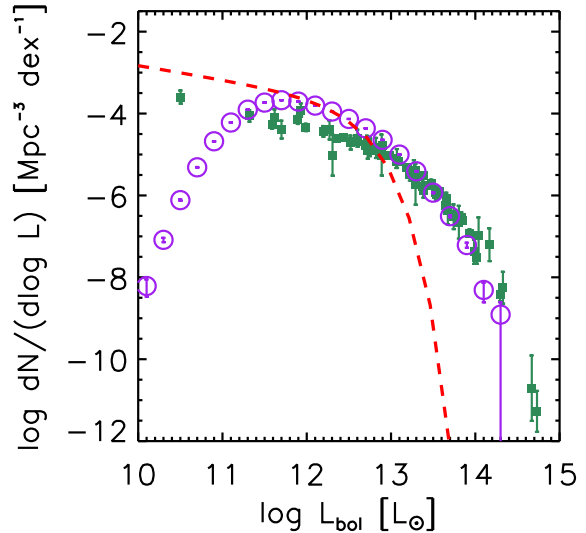}
\end{center}
\caption{\small The QSO LF at $z=2$ as shown in
  Fig.~\protect\ref{fig:bhev_s0.3}, with $\Gamma=1$ and
  $\sigmabh=0.3$, but with a QSO duty cycle of $\fagn=0.3$ and an
  Eddington ratio of $\fedd=0.5$.
\label{fig:bhev_g1_fedd}}
\end{figure}

One can try to sharpen this constraint by adopting more physically
reasonable values for the duty cycles and Eddington ratios of
AGN. \citet{erb:06} and \citet{kriek:07} find that about 20--40\% of
galaxies at $z\sim2$ contain an active nucleus, and models in which
such activity is merger-driven
\citep[e.g.][]{hopkins:07,somerville:08} predict that this fraction is
nearly constant for galaxy masses $10.0 \lesssim \log (m_*/\msun)
\lesssim 12.0$ \citep[see e.g. Figure 19
  of][]{hopkins:07}. \citet{vestergaard:04} find that the Eddington
ratios of luminous quasars at $1.5 < z < 3.5$ are in the range $0.1 <
L/L_{\rm Edd} < 1$, with an average value $L/L_{\rm Edd} \sim
0.4$--0.5, while \citet{kollmeier:06} find a fairly sharply peaked
distribution of Eddington ratios with a peak at $L/L_{\rm
  Edd}=0.25$. Adopting average values for the fraction of galaxies
containing active BH, $\fagn=0.3$, and the Eddington ratio $\fedd
\equiv L/L_{\rm Edd} =0.5$ (assuming that $L/L_{\rm Edd}$ is also
constant with galaxy/BH mass)\footnote{Note that if we included a
  realistic distribution of Eddington ratios, rather than a single
  constant value, this would again broaden the tail of the bright end
  of the QSO LF, leading to more very luminous QSOs.} produces quite
good agreement with the observed QSO LF at $z=2$ with {\em no
  evolution} in the zero-point or scatter of the \mgal-\mbh\ relation
($\Gamma=1$, $\sigma_{BH} = 0.3$; see Fig.~\ref{fig:bhev_g1_fedd}).

\begin{figure*} 
\begin{center}
\plottwo{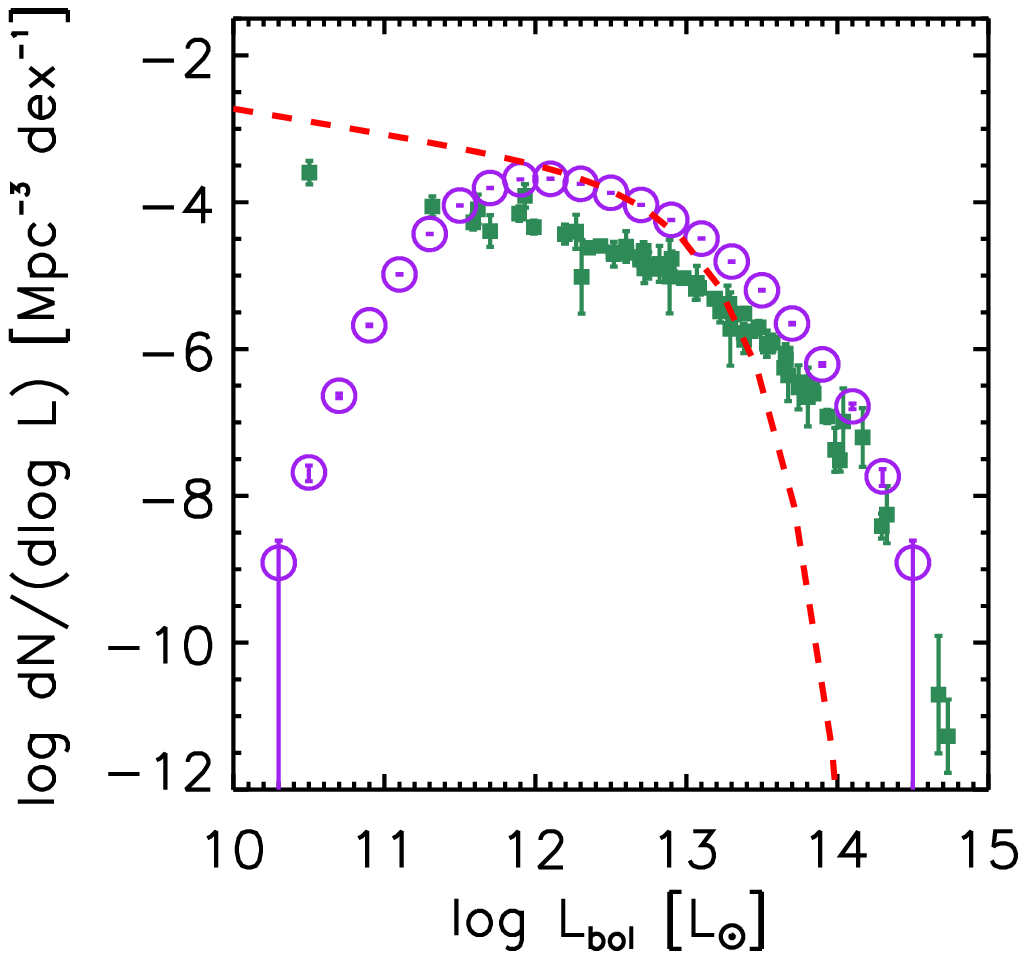}{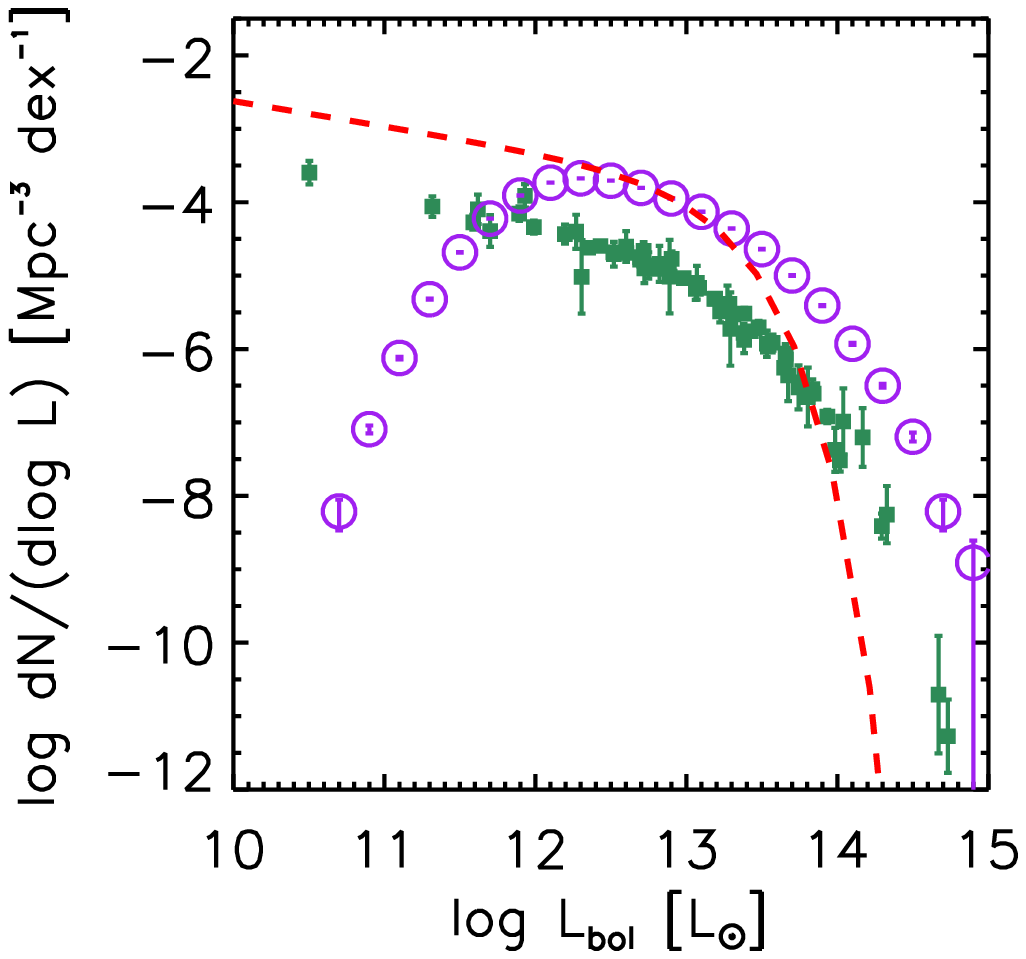}
\end{center}
\caption{\small The QSO LF at $z=2$ as shown in
  Fig.~\protect\ref{fig:bhev_s0.3}, with intrinsic scatter
  $\sigmabh=0.3$, a QSO duty cycle of $\fagn=0.3$ and an Eddington
  ratio of $\fedd=0.5$. Left panel: $\Gamma=2$; Right panel:
  $\Gamma=5$. Assuming that the duty cycles and Eddington ratios
  derived from independent observations are correct, and that the
  intrinsic scatter in the \mgal-\mbh\ relation was at least as large
  at $z=2$ as it is today, large amounts of evolution in the
  zero-point ($\Gamma \gtrsim 2$) are disfavored. 
\label{fig:bhev_g2_fedd}}
\end{figure*}

Turning the argument around, then, if the independent observational
estimates of duty cycle and Eddington ratio are correct, and if the
scatter in the \mgal-\mbh\ relation was not significantly smaller at
high redshift than it is today, then overall evolution in the
\mgal-\mbh\ relation of $\Gamma \gtrsim 2$ since $z\sim2$ is
disfavored as it would {\em overproduce} the number of luminous QSOs
(see Fig.~\ref{fig:bhev_g2_fedd}). In particular, if the value of
$\Gamma$ at $z\sim2$ were as large as suggested by the observations of
e.g. \citet[][]{peng:06}, $\Gamma \gtrsim 4$, the number of luminous
QSOs would be overproduced by more than one order of magnitude. Of
course, one could reconcile these larger amounts of evolution if the
duty cycle of luminous quasars is an order of magnitude smaller than
what I have assumed ($\sim 2$--3 percent instead of $20$--30
percent).

%=======================
% 3
\section{Conclusions}
\label{sec:conclusions}
%=======================

I have investigated whether observational estimates of the stellar
mass function of galaxies, combined with observed QSO luminosity
functions, can provide useful limits on the relationship between
galaxies and their SMBH at high redshift. I assumed a simple
relationship between galaxy mass and SMBH mass, as observed in dormant
galaxies in the nearby Universe, and a simple form for the possible
evolution of this relationship (see Eqn.~\ref{eqn:mbhev}), namely a
shift in the zero-point of the relation by a redshift-dependent factor
$\Gamma(z)$. I then argued that one can obtain a {\em lower} limit on
$\Gamma(z)$ by making the rather extreme assumption that all BH
radiate at their Eddington limit at all times, and requiring that at
least the observed number of luminous QSOs be reproduced. I further
argued that an {\em upper} limit on $\Gamma(z)$ could be obtained by
requiring that the number of massive BH in galaxies today should not
be exceeded at high redshift.

Assuming that there is a deterministic relationship between galaxy
mass and BH mass (i.e., no scatter in the \mbh-\mgal\ relationship), I
find that in order to produce enough luminous QSOs, the zero-point of
the relation must have been higher by at least a factor of $\sim 2$ at
$z=1$ and a factor of 5--6 at $z=2$. At the same time, in order to
avoid producing a larger number density of massive BH than what is
implied by observations at $z\sim0$, the upper limit on the evolution
of the normalization of the \mbh-\mgal\ relationship at $z=2$ is about
a factor of six. Since both the lower and upper limits are fairly
liberal, one might have expected them to lie several orders of
magnitude apart, and therefore not to provide very interesting
constraints on the actual evolution of the
\mbh-\mgal\ relationship. It seems potentially quite interesting that
these limits lie nearly on top of one another.

However, relaxing the assumption of a perfectly deterministic
\mbh-\mgal\ relationship has a major impact on the results. When
scatter is included in the \mbh-\mgal\ relation at a level similar to
the intrinsic scatter in the observed relation at $z=0$, I find that
the majority of very massive BH are objects that live in galaxies of
moderate mass but are outliers in the \mbh-\mgal\ relationship. This
is of course due to the very steep slope of the galaxy stellar mass
function at large masses. Because the constraints above arose from the
most luminous QSOs, I then find that there is a strong degeneracy
between the evolution of the zero-point $\Gamma(z)$ and the scatter
$\sigmabh$. For example, the QSO constraint at $z=2$ can be reproduced
even in a scenario in which $\Gamma=1$ (no evolution in the zero-point
has occured) and $\sigmabh=0.3$ (the intrinsic scatter in
\mbh-\mgal\ is similar to that in the observed relation today). Thus
we are left with the very weak constraint that BH probably were no
{\em smaller} at high redshift relative to their host galaxies (unless
the scatter was much larger than it is today).

I tried to sharpen this constraint by adopting more physically
reasonable values for the duty cycles and Eddington ratios of AGN,
based on independent observational constraints. Adopting
mass-independent values of $\fagn=0.3$ (the fraction of galaxies
hosting AGN) and $\fedd \equiv L/L_{\rm Edd} \sim 0.5$, and assuming a
scatter in \mbh-\mgal\ similar to that in the observed relation for
dormant galaxies today ($\sigmabh=0.3$), I find that BH cannot have
been much more than a factor of $\sim 2$ more massive relative to
their host galaxies at $z\sim 2$ than they are today. In particular,
values as large as $\Gamma(z=2) \sim 4$, as suggested by some
observational studies \citep[e.g.][]{peng:06}, would overproduce the
number of luminous QSOs by more than an order of magnitude.

Interestingly, \citet{hopkins:06} also reached similar conclusions
based on a somewhat different, though related argument. They pointed
out that in order to avoid overproducing the {\em total} mass density
in SMBH relative to the present day value, the average value of
\mbh/\mgal\ must not have been more than about a factor of two larger
at $z\sim2$ than today's value.

I have based these results on the relationship between the total
stellar mass of the galaxy and the mass of the SMBH; however, there is
strong evidence that the more fundamental relationship is actually
between the BH mass and the mass of the {\em spheroidal} component of
the galaxy \citep[e.g.][]{kormendy_review:95}. I made this choice
because the stellar mass function of galactic spheroids is very poorly
constrained at high redshift. However, at low redshift, the most
massive galaxies are predominantly spheroid-dominated
\citep[e.g.][]{bell:03}. If this was also the case at high redshift,
then my conclusions will not change much as the constraints are driven
by the most massive BH which are hosted by massive galaxies. If there
is a significant population of disk-dominated massive galaxies at high
redshift, and the BH mass indeed correlates with spheroid mass only,
then this would leave more room for evolution and/or scatter in the
\msph-\mbh\ relation.

Another source of uncertainty arises from the fact that BH masses
predicted from the \mbh\ vs. luminosity (\mbh-$L$) relationship are
inconsistent with those predicted from the \mbh\ vs. velocity
dispersion (\mbh-$\sigma$) relationship for the most luminous galaxies
\citep{lauer:07a}. The \mbh-\mgal\ relationship that I have chosen to
use here is derived from the \mbh-$L$ relation, which
\citet{lauer:07a} argue should be more reliable in the regime of
interest, but the situation at high redshift is unknown. Currently,
there are no published observational measurements of the galaxy
velocity dispersion function at high redshift (of which I am aware);
however, these may become available in the future. It would then be
very interesting to repeat this kind of analysis using \mbh-$\sigma$
instead.

Although it is dissappointing that the proposed approach did not yield
stronger constraints on the evolution of the \mbh-\mgal\ relationship,
this exercise has brought out a few important lessons. First, in order
to understand the relationship between galaxies and their BH, it is
perhaps as important to understand the magnitude and evolution of the
{\em intrinsic scatter} in this relationship as it is to understand
the evolution of the zero-point of the relation itself. Second, new
generations of theoretical models that attempt to simultaneously treat
the formation and evolution of galaxies and their black holes
\citep[e.g.][]{croton:06,bower:06,fontanot:06,somerville:08} must take
care to properly model the dispersion in the \mbh-\mgal\ relationship.

%===================================
\section*{Acknowledgments}
\begin{small}

I thank Sandra Faber, Cole Miller, Brant Robertson, Tod Lauer, and
Chien Peng for useful comments on this work, and Phil Hopkins, Lars
Hernquist, T.J. Cox, Yuexing Li, and Brant Robertson for stimulating
discussions. I also thank the anonymous referee for comments that
improved the paper.

\end{small}
%=====================================

\bibliographystyle{mn} 
\bibliography{mn-jour,bhev}

\begin{thebibliography}{55}
\expandafter\ifx\csname natexlab\endcsname\relax\def\natexlab#1{#1}\fi

\bibitem[{{Adams} {et~al.}(2003){Adams}, {Graff}, {Mbonye}, \&
  {Richstone}}]{adams:03}
{Adams} F.~C., {Graff} D.~S., {Mbonye} M., {Richstone} D.~O., 2003, \apj, 591,
  125

\bibitem[{{Adams} {et~al.}(2001){Adams}, {Graff}, \& {Richstone}}]{adams:01}
{Adams} F.~C., {Graff} D.~S., {Richstone} D.~O., 2001, \apjl, 551, L31

\bibitem[{{Bell} \& {de Jong}(2001)}]{bell_dejong:01}
{Bell} E.~F., {de Jong} R.~S., 2001, \apj, 550, 212

\bibitem[{{Bell} {et~al.}(2003){Bell}, {McIntosh}, {Katz}, \&
  {Weinberg}}]{bell:03}
{Bell} E.~F., {McIntosh} D.~H., {Katz} N., {Weinberg} M.~D., 2003, \apjs, 149,
  289

\bibitem[{{Borch} {et~al.}(2006){Borch}, {Meisenheimer}, {Bell}, {Rix}, {Wolf},
  {Dye}, {Kleinheinrich}, {Kovacs}, \& {Wisotzki}}]{Borch:06}
{Borch} A., {Meisenheimer} K., {Bell} E.~F., {Rix} H.-W., {Wolf} C., {Dye} S.,
  {Kleinheinrich} M., {Kovacs} Z., {Wisotzki} L., 2006, \aap, 453, 869

\bibitem[{{Bower} {et~al.}(2006){Bower}, {Benson}, {Malbon}, {Helly}, {Frenk},
  {Baugh}, {Cole}, \& {Lacey}}]{bower:06}
{Bower} R.~G., {Benson} A.~J., {Malbon} R., {Helly} J.~C., {Frenk} C.~S.,
  {Baugh} C.~M., {Cole} S., {Lacey} C.~G., 2006, \mnras, 370, 645

\bibitem[{{Bundy} {et~al.}(2006){Bundy}, {Ellis}, {Conselice}, {Taylor},
  {Cooper}, {Willmer}, {Weiner}, {Coil}, {Noeske}, \& {Eisenhardt}}]{Bundy:06}
{Bundy} K., {Ellis} R.~S., {Conselice} C.~J., {Taylor} J.~E., {Cooper} M.~C.,
  {Willmer} C.~N.~A., {Weiner} B.~J., {Coil} A.~L., {Noeske} K.~G.,
  {Eisenhardt} P.~R.~M., 2006, \apj, 651, 120

\bibitem[{{Burkert} \& {Silk}(2001)}]{burkert_silk:01}
{Burkert} A., {Silk} J., 2001, \apjl, 554, L151

\bibitem[{{Chabrier}(2003)}]{chabrier:03}
{Chabrier} G., 2003, \apjl, 586, L133

\bibitem[{{Croton}(2006)}]{croton:bhev}
{Croton} D.~J., 2006, \mnras, 369, 1808

\bibitem[{{Croton} {et~al.}(2006){Croton}, {Springel}, {White}, {De Lucia},
  {Frenk}, {Gao}, {Jenkins}, {Kauffmann}, {et~al.}}]{croton:06}
{Croton} D.~J., {Springel} V., {White} S.~D.~M., {De Lucia} G., {Frenk} C.~S.,
  {Gao} L., {Jenkins} A., {Kauffmann} G., {et~al.}, 2006, \mnras, 365, 11

\bibitem[{{Dressler}(1989)}]{dressler:89}
{Dressler} A., 1989, in IAU Symposium, Vol. 134, Active Galactic Nuclei,
  {Osterbrock} D.~E., {Miller} J.~S., eds., pp. 217--+

\bibitem[{{Drory} {et~al.}(2004){Drory}, {Bender}, {Feulner}, {Hopp},
  {Maraston}, {Snigula}, \& {Hill}}]{Drory:04}
{Drory} N., {Bender} R., {Feulner} G., {Hopp} U., {Maraston} C., {Snigula} J.,
  {Hill} G.~J., 2004, \apj, 608, 742

\bibitem[{{Erb} {et~al.}(2006){Erb}, {Steidel}, {Shapley}, {Pettini}, {Reddy},
  \& {Adelberger}}]{erb:06}
{Erb} D.~K., {Steidel} C.~C., {Shapley} A.~E., {Pettini} M., {Reddy} N.~A.,
  {Adelberger} K.~L., 2006, \apj, 646, 107

\bibitem[{{Ferrarese} \& {Ford}(2005)}]{ferrarese_ford:05}
{Ferrarese} L., {Ford} H., 2005, Space Science Reviews, 116, 523

\bibitem[{{Ferrarese} \& {Merritt}(2000)}]{ferrarese:00}
{Ferrarese} L., {Merritt} D., 2000, \apjl, 539, L9

\bibitem[{{Fontana} {et~al.}(2004){Fontana}, {Pozzetti}, {Donnarumma},
  {Renzini}, {Cimatti}, {Zamorani}, {Menci}, {Daddi}, {Giallongo}, {Mignoli},
  {Perna}, {Salimbeni}, {Saracco}, {Broadhurst}, {Cristiani}, {D'Odorico}, \&
  {Gilmozzi}}]{Fontana:04}
{Fontana} A., {Pozzetti} L., {Donnarumma} I., {Renzini} A., {Cimatti} A.,
  {Zamorani} G., {Menci} N., {Daddi} E., {Giallongo} E., {Mignoli} M., {Perna}
  C., {Salimbeni} S., {Saracco} P., {Broadhurst} T., {Cristiani} S.,
  {D'Odorico} S., {Gilmozzi} R., 2004, \aap, 424, 23

\bibitem[{{Fontana} {et~al.}(2006){Fontana}, {Salimbeni}, {Grazian},
  {Giallongo}, {Pentericci}, {Nonino}, {Fontanot}, {Menci}, {Monaco},
  {Cristiani}, {Vanzella}, {de Santis}, \& {Gallozzi}}]{Fontana:06}
{Fontana} A., {Salimbeni} S., {Grazian} A., {Giallongo} E., {Pentericci} L.,
  {Nonino} M., {Fontanot} F., {Menci} N., {Monaco} P., {Cristiani} S.,
  {Vanzella} E., {de Santis} C., {Gallozzi} S., 2006, \aap, 459, 745

\bibitem[{{Fontanot} {et~al.}(2009){Fontanot}, {De Lucia}, {Monaco},
  {Somerville}, \& {Santini}}]{fontanot:09}
{Fontanot} F., {De Lucia} G., {Monaco} P., {Somerville} R.~S., {Santini} P.,
  2009, astro-ph/0901.1130

\bibitem[{{Fontanot} {et~al.}(2006){Fontanot}, {Monaco}, {Cristiani}, \&
  {Tozzi}}]{fontanot:06}
{Fontanot} F., {Monaco} P., {Cristiani} S., {Tozzi} P., 2006, \mnras, 373, 1173

\bibitem[{{Gebhardt} {et~al.}(2000)}]{gebhardt:00}
{Gebhardt} K., {et~al.}, 2000, \apjl, 539, L13

\bibitem[{{Graham} {et~al.}(2001){Graham}, {Erwin}, {Caon}, \&
  {Trujillo}}]{graham:01}
{Graham} A.~W., {Erwin} P., {Caon} N., {Trujillo} I., 2001, \apjl, 563, L11

\bibitem[{{Gultekin} {et~al.}(2009){Gultekin}, {Richstone}, {Gebhardt},
  {Lauer}, {Tremaine}, {Aller}, {Bender}, {Dressler}, {Faber}, {Filippenko},
  {Green}, {Ho}, {Kormendy}, {Magorrian}, {Pinkney}, \& {Siopis}}]{gultekin:09}
{Gultekin} K., {Richstone} D.~O., {Gebhardt} K., {Lauer} T.~R., {Tremaine} S.,
  {Aller} M.~C., {Bender} R., {Dressler} A., {Faber} S.~M., {Filippenko} A.~V.,
  {Green} R., {Ho} L.~C., {Kormendy} J., {Magorrian} J., {Pinkney} J., {Siopis}
  C., 2009, astro-ph/0903.4897

\bibitem[{{H{\"a}ring} \& {Rix}(2004)}]{haering:04}
{H{\"a}ring} N., {Rix} H.-W., 2004, \apjl, 604, L89

\bibitem[{{Hopkins} {et~al.}(2008){Hopkins}, {Hernquist}, {Cox}, \& {Kere{\v
  s}}}]{hopkins:07}
{Hopkins} P.~F., {Hernquist} L., {Cox} T.~J., {Kere{\v s}} D., 2008, \apjs,
  175, 356

\bibitem[{{Hopkins} {et~al.}(2007{\natexlab{a}}){Hopkins}, {Hernquist}, {Cox},
  {Robertson}, \& {Krause}}]{hopkins_bhfpth:07}
{Hopkins} P.~F., {Hernquist} L., {Cox} T.~J., {Robertson} B., {Krause} E.,
  2007{\natexlab{a}}, \apj, 669, 45

\bibitem[{{Hopkins} {et~al.}(2007{\natexlab{b}}){Hopkins}, {Richards}, \&
  {Hernquist}}]{hopkins_qsolf:07}
{Hopkins} P.~F., {Richards} G.~T., {Hernquist} L., 2007{\natexlab{b}}, \apj,
  654, 731

\bibitem[{{Hopkins} {et~al.}(2006){Hopkins}, {Robertson}, {Krause},
  {Hernquist}, \& {Cox}}]{hopkins:06}
{Hopkins} P.~F., {Robertson} B., {Krause} E., {Hernquist} L., {Cox} T.~J.,
  2006, \apj, 652, 107

\bibitem[{{Kauffmann} {et~al.}(2003)}]{kauffmann:03}
{Kauffmann} G., {et~al.}, 2003, \mnras, 341, 54

\bibitem[{{Kollmeier} {et~al.}(2006)}]{kollmeier:06}
{Kollmeier} J.~A., {et~al.}, 2006, \apj, 648, 128

\bibitem[{{Kormendy} \& {Bender}(2009)}]{kormendy:09}
{Kormendy} J., {Bender} R., 2009, \apjl, 691, L142

\bibitem[{{Kormendy} \& {Richstone}(1995)}]{kormendy_review:95}
{Kormendy} J., {Richstone} D., 1995, \araa, 33, 581

\bibitem[{{Kriek} {et~al.}(2007){Kriek}, {van Dokkum}, {Franx}, {Illingworth},
  {Coppi}, {F{\"o}rster Schreiber}, {Gawiser}, {Labb{\'e}}, {Lira},
  {Marchesini}, {Quadri}, {Rudnick}, {Taylor}, {Urry}, \& {van der
  Werf}}]{kriek:07}
{Kriek} M., {van Dokkum} P.~G., {Franx} M., {Illingworth} G.~D., {Coppi} P.,
  {F{\"o}rster Schreiber} N.~M., {Gawiser} E., {Labb{\'e}} I., {Lira} P.,
  {Marchesini} D., {Quadri} R., {Rudnick} G., {Taylor} E.~N., {Urry} C.~M.,
  {van der Werf} P.~P., 2007, \apj, 669, 776

\bibitem[{{Lauer} {et~al.}(2007{\natexlab{a}}){Lauer}, {Tremaine}, {Richstone},
  \& {Faber}}]{lauer:07}
{Lauer} T.~R., {Tremaine} S., {Richstone} D., {Faber} S.~M.,
  2007{\natexlab{a}}, \apj, 670, 249

\bibitem[{{Lauer} {et~al.}(2007{\natexlab{b}})}]{lauer:07a}
{Lauer} T.~R., {et~al.}, 2007{\natexlab{b}}, \apj, 662, 808

\bibitem[{{Magorrian} {et~al.}(1998)}]{magorrian:98}
{Magorrian} J., {et~al.}, 1998, \aj, 115, 2285

\bibitem[{{Marchesini} {et~al.}(2008){Marchesini}, {van Dokkum}, {Forster
  Schreiber}, {Franx}, {Labbe'}, \& {Wuyts}}]{Marchesini:08}
{Marchesini} D., {van Dokkum} P.~G., {Forster Schreiber} N.~M., {Franx} M.,
  {Labbe'} I., {Wuyts} S., 2008, astro-ph/0811.1773

\bibitem[{{Marconi} \& {Hunt}(2003)}]{marconi:03}
{Marconi} A., {Hunt} L.~K., 2003, \apjl, 589, L21

\bibitem[{{Marconi} {et~al.}(2004){Marconi}, {Risaliti}, {Gilli}, {Hunt},
  {Maiolino}, \& {Salvati}}]{marconi:04}
{Marconi} A., {Risaliti} G., {Gilli} R., {Hunt} L.~K., {Maiolino} R., {Salvati}
  M., 2004, \mnras, 351, 169

\bibitem[{{Novak} {et~al.}(2006){Novak}, {Faber}, \& {Dekel}}]{novak:06}
{Novak} G.~S., {Faber} S.~M., {Dekel} A., 2006, \apj, 637, 96

\bibitem[{{Pannella} {et~al.}(2006){Pannella}, {Hopp}, {Saglia}, {Bender},
  {Drory}, {Salvato}, {Gabasch}, \& {Feulner}}]{Pannella:06}
{Pannella} M., {Hopp} U., {Saglia} R.~P., {Bender} R., {Drory} N., {Salvato}
  M., {Gabasch} A., {Feulner} G., 2006, \apjl, 639, L1

\bibitem[{{Peng} {et~al.}(2006){Peng}, {Impey}, {Rix}, {Kochanek}, {Keeton},
  {Falco}, {Leh{\'a}r}, \& {McLeod}}]{peng:06}
{Peng} C.~Y., {Impey} C.~D., {Rix} H.-W., {Kochanek} C.~S., {Keeton} C.~R.,
  {Falco} E.~E., {Leh{\'a}r} J., {McLeod} B.~A., 2006, \apj, 649, 616

\bibitem[{{P{\'e}rez-Gonz{\'a}lez} {et~al.}(2008){P{\'e}rez-Gonz{\'a}lez},
  {Rieke}, {Villar}, {Barro}, {Blaylock}, {Egami}, {Gallego}, {Gil de Paz},
  {Pascual}, {Zamorano}, \& {Donley}}]{PerezGonzalez:08}
{P{\'e}rez-Gonz{\'a}lez} P.~G., {Rieke} G.~H., {Villar} V., {Barro} G.,
  {Blaylock} M., {Egami} E., {Gallego} J., {Gil de Paz} A., {Pascual} S.,
  {Zamorano} J., {Donley} J.~L., 2008, \apj, 675, 234

\bibitem[{{Pozzetti} {et~al.}(2007){Pozzetti}, {Bolzonella}, {Lamareille},
  {Zamorani}, {Franzetti}, {Le F{\`e}vre}, {Iovino}, \&
  {Temporin}}]{Pozzetti:07}
{Pozzetti} L., {Bolzonella} M., {Lamareille} F., {Zamorani} G., {Franzetti} P.,
  {Le F{\`e}vre} O., {Iovino} A., {Temporin} S. e.~a., 2007, \aap, 474, 443

\bibitem[{{Robertson} {et~al.}(2006){Robertson}, {Hernquist}, {Cox}, {Di
  Matteo}, {Hopkins}, {Martini}, \& {Springel}}]{robertson:06}
{Robertson} B., {Hernquist} L., {Cox} T.~J., {Di Matteo} T., {Hopkins} P.~F.,
  {Martini} P., {Springel} V., 2006, \apj, 641, 90

\bibitem[{{Salviander} {et~al.}(2007){Salviander}, {Shields}, {Gebhardt}, \&
  {Bonning}}]{salviander:07}
{Salviander} S., {Shields} G.~A., {Gebhardt} K., {Bonning} E.~W., 2007, \apj,
  662, 131

\bibitem[{{Silk} \& {Rees}(1998)}]{silk_rees:98}
{Silk} J., {Rees} M.~J., 1998, \aap, 331, L1

\bibitem[{{Somerville} {et~al.}(2008){Somerville}, {Hopkins}, {Cox},
  {Robertson}, \& {Hernquist}}]{somerville:08}
{Somerville} R.~S., {Hopkins} P.~F., {Cox} T.~J., {Robertson} B.~E.,
  {Hernquist} L., 2008, \mnras, 391, 481

\bibitem[{{Tremaine} {et~al.}(2002)}]{tremaine:02}
{Tremaine} S., {et~al.}, 2002, \apj, 574, 740

\bibitem[{{Treu} {et~al.}(2004){Treu}, {Malkan}, \& {Blandford}}]{treu:04}
{Treu} T., {Malkan} M.~A., {Blandford} R.~D., 2004, \apjl, 615, L97

\bibitem[{{Vergani} {et~al.}(2008){Vergani}, {Scodeggio}, {Pozzetti}, {Iovino},
  {Franzetti}, {Garilli}, {Zamorani}, \& {Maccagni}}]{Vergani:08}
{Vergani} D., {Scodeggio} M., {Pozzetti} L., {Iovino} A., {Franzetti} P.,
  {Garilli} B., {Zamorani} G., {Maccagni} D. e.~a., 2008, \aap, 487, 89

\bibitem[{{Vestergaard}(2004)}]{vestergaard:04}
{Vestergaard} M., 2004, \apj, 601, 676

\bibitem[{{Woo} {et~al.}(2006){Woo}, {Treu}, {Malkan}, \& {Blandford}}]{woo:06}
{Woo} J.-H., {Treu} T., {Malkan} M.~A., {Blandford} R.~D., 2006, \apj, 645, 900

\bibitem[{{Woo} {et~al.}(2008){Woo}, {Treu}, {Malkan}, \& {Blandford}}]{woo:08}
---, 2008, \apj, 681, 925

\bibitem[{{Wyithe} \& {Loeb}(2003)}]{wyithe_loeb:03}
{Wyithe} J.~S.~B., {Loeb} A., 2003, \apj, 595, 614

\end{thebibliography}

\end{document}